\documentclass[aps,prl,reprint,groupedaddress,showpacs]{revtex4-1}
\usepackage{graphicx}
\usepackage{amsmath}
\usepackage{amsfonts}
\usepackage{amssymb}
\usepackage{wasysym}
\usepackage{float}

\newcommand{\omegabold}{\boldsymbol\omega}

\begin{document}

\title{Development of high vorticity in incompressible 3D Euler equations: influence of initial conditions}
\thanks{dmitrij@itp.ac.ru}

\author{D.S. Agafontsev$^{(a),(b)}$, E.A. Kuznetsov$^{(b),(c)}$ and A.A.
Mailybaev$^{(d)}$}
\affiliation{$^{(a)}$ P. P. Shirshov Institute of Oceanology, Moscow, Russia\\
	$^{(b)}$ Novosibirsk State University, Novosibirsk, Russia\\
	$^{(c)}$ P.N. Lebedev Physical Institute, Moscow, Russia\\
	$^{(d)}$ Instituto Nacional de Matem\'atica Pura e Aplicada -- IMPA, Rio de Janeiro, Brazil}

\begin{abstract}
The incompressible three-dimensional ideal flows develop very thin pancake-like regions of increasing vorticity. 
These regions evolve with the scaling $\omega_{\max}(t)\propto\ell(t)^{-2/3}$ between the vorticity maximum and pancake thickness, and provide the leading contribution to the energy spectrum, where the gradual formation of the Kolmogorov interval $E_{k}\propto k^{-5/3}$ is observed for some initial flows \textit{[Agafontsev et. al, Phys. Fluids 27, 085102 (2015)]}. 
With the massive numerical simulations, in the present paper we study the influence of initial conditions on the processes of pancake formation and the Kolmogorov energy spectrum development. 
\end{abstract}

\maketitle


\section{Introduction}

According to the Kolmogorov-Obukhov theory of developed hydrodynamic turbulence~\cite{kolmogorov1941local,obukhov1941spectral}, the velocity fluctuations at intermediate spatial scales $l$ obey the power-law $\langle|\delta v|\rangle\propto \varepsilon^{1/3} l^{1/3}$, where $\varepsilon$ is the mean energy flux from large to small scales. 
Consequently, fluctuations of the vorticity field $\omegabold=\mathrm{rot}\,\mathbf{v}$ diverge at small scales as $\langle|\delta\omega|\rangle\propto \varepsilon^{1/3} l^{-2/3}$. 
These relations lead to $E_{k}\propto\varepsilon^{2/3} k^{-5/3}$ for the energy spectrum in the corresponding range of wavenumbers $k$ (the inertial interval), see e.g.~\cite{landau2013fluid,frisch1999turbulence}.
This implies that the Kolmogorov spectrum is linked with the small-scale structures of intense vorticity. 
The Kolmogorov arguments are based on the isotropy of the flow and locality of nonlinear interaction at intermediate scales. 
Then, the dynamics in the inertial interval can be described by the Euler equations and the emergence of the Kolmogorov spectrum can be expected before the viscous scales get excited~\cite{orlandipirozzoli2010,holm2002transient,cichowlas2005effective,holm2007}.  

In our previous paper~\cite{agafontsev2015}, we studied the development of high-vorticity regions in incompressible 3D Euler equations with high-accuracy numerical simulations for two initial conditions. 
According to~\cite{brachet1992numerical,frisch2003singularities}, in the generic case such regions represent exponentially compressing pancake-like structures (thin vorticity sheets). 
Contrary to these two studies, we found that evolution of the pancakes is governed by two different exponents $\ell(t)\propto e^{-t/T_{\ell}}$ and $\omega_{\max}(t)\propto e^{t/T_{\omega}}$ describing compression in the transverse direction and vorticity growth respectively, with the universal ratio $T_{\ell}/T_{\omega}\approx 2/3$. 
This ratio leads to the Kolmogorov-type scaling $\omega_{\max}(t)\propto\ell(t)^{-2/3}$ between the vorticity maximum and pancake thickness. 
The pancakes appear in increasing number with different scales. 
We demonstrated that these structures generate strongly anisotropic vorticity field in the Fourier space, concentrated in ``jets'' extended in the directions perpendicular to the pancakes. 
These jets, occupying only a small fraction of the entire spectral space, dominate in the energy spectrum, where we observed clearly formation of the interval with the Kolmogorov scaling $E_{k}\propto k^{-5/3}$, in a fully inviscid flow. 

In the present paper we continue this study, examining the influence of initial conditions on the processes of pancake formation and the Kolmogorov energy spectrum development. 
We perform simulations for $30$ initial conditions in the form of a superposition of the shear flow $\omega_{x}=\sin z$, $\omega_{y}=\cos z$, $\omega_{z}=0$ and a random (not necessarily small) perturbation. 
The presence of the shear flow influences the orientation of emerging pancake structures, from fully random when the shear flow is absent to almost unidirectional close to $z$-axis when the perturbation is small. 
We observe that the scaling $\omega_{\max}(t)\propto\ell(t)^{-2/3}$ between the vorticity maximum and pancake thickness holds universally, while initial conditions composed of the shear flow and a small perturbation develop significantly longer intervals of wavenumbers with the energy spectrum close to the Kolmogorov law $E_{k}\propto k^{-5/3}$. 


\section{Numerical methods}

We integrate the incompressible 3D Euler equations (in the vorticity formulation) 
\begin{equation}\label{Euler2}
\frac{\partial\omegabold}{\partial t} = \mathrm{rot}\,(\mathbf{v}\times \omegabold),\quad
\mathbf{v} = \mathrm{rot}^{-1}\omegabold,
\end{equation}
in the periodic box $\mathbf{r}=(x,y,z)\in \lbrack -\pi ,\pi ]^{3}$ using pseudo-spectral Runge-Kutta fourth-order method.  
To avoid the so-called bottle-neck instability, we use filtering in the Fourier space at each time step with the cut-off function suggested in~\cite{hou2007computing},
\begin{eqnarray}\label{HLdumping}
\rho(\mathbf{k}) = \exp\bigg(-36\sum_{j=x,y,z}(k_{j}/K_{\max}^{(j)})^{36}\bigg),
\end{eqnarray}
where $\mathbf{k} = (k_{x},k_{y},k_{z})$ is wavevector, $K_{\max}^{(j)}=N_{j}/2$ are the maximal wavenumbers and $N_{j}$ are numbers of nodes along directions $j=x,y,z$, so that $|k_{j}|\le K_{\max}^{(j)}$. 
Function~(\ref{HLdumping}) cuts off approximately 20\% of the spectrum at the edges of the spectral band in each direction. 
The inverse of the curl operator in Eq.~(\ref{Euler2}) is calculated in the Fourier space, see, e.g.,~\cite{agafontsev2015}. 
Adaptive time stepping is implemented through the CFL stability criterion with the Courant number $0.5$. 

We use adaptive anisotropic rectangular grid, which is uniform in each direction and adapted independently along each of the three coordinates. 
The idea for the adaption comes from the standard dealiasing rule, optimized for the quadratic nonlinearity of the Euler equations. 
At early times, the Fourier spectrum of the solution is concentrated at low harmonics, while higher harmonics contain numerical noise. 
We track the ``signal-noise'' boundary~\cite{agafontsev2015} until it reaches $2K_{\max }^{(j)}/3$ for any of the three directions $j=x,y,z$. 
Then we refine the grid along the corresponding direction using the Fourier interpolation, which has an error comparable with the round-off. 
While the simulation is running in this way, the aliasing error is avoided and the influence of the filtering~(\ref{HLdumping}) on the ``signal-containing'' harmonics is negligible.
We start simulations in cubic grid $128^{3}$. 
When the total number of nodes reaches $1024^{3}$, we continue with the fixed grid until the Fourier spectrum of the vorticity at $2K_{\max }^{(j)}/3$ exceeds $10^{-10}$ times its maximum value, see~\cite{agafontsev2015}, along any of the three directions $j$. 

We perform simulations of $30$ initial flows taken as superpositions of the shear flow 
\begin{equation}\label{IC}
\omegabold_{sh}(\mathbf{r}) = (\sin z, \cos z, 0),\quad |\omegabold_{sh}(\mathbf{r})|=1,
\end{equation}
and a random periodic perturbation (different for each initial flow) 
\begin{equation}
\omegabold_{p}(\mathbf{r}) =  \sum_{\mathbf{h}} 
\left[\mathbf{A}_\mathbf{h}\cos(\mathbf{h}\cdot\mathbf{r})
+\mathbf{B}_\mathbf{h}\sin(\mathbf{h}\cdot\mathbf{r})\right].
\label{IC1}
\end{equation}
Here $\mathbf{h} = (h_x,h_y,h_z)$ is a vector with integer components $|h_{j}|\le 2$, $j=x,y,z$, while vectors $\mathbf{A}_\mathbf{h}$ and $\mathbf{B}_\mathbf{h}$ of real random coefficients of zero mean and variance $\sigma_{\mathbf{h}}^2 \sim \exp(-|\mathbf{h}|^2)$ satisfy the orthogonality conditions, $\mathbf{h}\cdot\mathbf{A}_\mathbf{h} = \mathbf{h}\cdot\mathbf{B}_\mathbf{h} = 0$, necessary for self-consistency. 
The shear flow~(\ref{IC}) is exact stationary solution of the Euler equations, with the velocity coinciding with vorticity $\mathbf{v}=\omegabold$. 

Three types of initial conditions are considered:
\begin{itemize}
 \item Type I -- generic periodic flows $\omegabold=\omegabold_{p}$;
 \item Type II -- mix of the shear flow with a random perturbation $\omegabold=\omegabold_{sh}+\omegabold_{p}$. Amplitude of the perturbation is adjusted, so that the mix satisfies $0.2\le |\omegabold|\le 1.8$, $\max|\omegabold|\ge 1.6$;
 \item Type III -- mix of the shear flow with a random perturbation $\omegabold=\omegabold_{sh}+\omegabold_{p}$. Amplitude of the perturbation is adjusted, so that the mix satisfies $0.8\le |\omegabold|\le 1.2$, $\max|\omegabold|\ge 1.15$. 
\end{itemize}
Thus, an initial condition of Type I is a fully generic periodic flow, Type II -- is a mix in proportions similar to $1:0.8$ of the shear flow with a random flow, and Type III -- is the shear flow with a small perturbation. 
For each type, we tested 10 random realizations of initial flows. 
The results are illustrated below for three representative simulations with initial conditions 
\begin{itemize}
 \item $IC_{1}$ (Type I), ended at the final time $t=6.52$ with the grid $1152\times 972\times 864$;
 \item $IC_{2}$ (Type II), ended at $t=7.72$ with the grid $768\times 864\times 1458$;
 \item $IC_{3}$ (Type III), ended at $t=12.57$ with the grid $432\times 864\times 2592$.
\end{itemize}

During the simulations, we track evolution of the energy spectrum 
\begin{equation}
E_{k}(t) = \frac{1}{2}\sum_{|\mathbf{p}|= k} |\mathbf{v}(\mathbf{p},t)|^{2}, \label{energy_spectrum}
\end{equation}
where $\mathbf{v}(\mathbf{p},t)$ is the Fourier-transformed velocity field and the integral sum is calculated by summation on spherical shells of unit width $\Delta k=1$. 
For some of the initial conditions we observe the gradual formation of the interval of wavenumbers, up to $2\lesssim k\lesssim 20$, with the scaling similar to the Kolmogorov one, $E_{k}\propto k^{-5/3}$. 
This interval is characterized by the ``frozen'' part of the spectrum, in contrast to the vast changes with time at larger wavenumbers, and contains only a small fraction of total energy (almost all energy is contained in the first harmonic $k=1$). 

\begin{figure}[t]
\centering
\includegraphics[width=8.5cm]{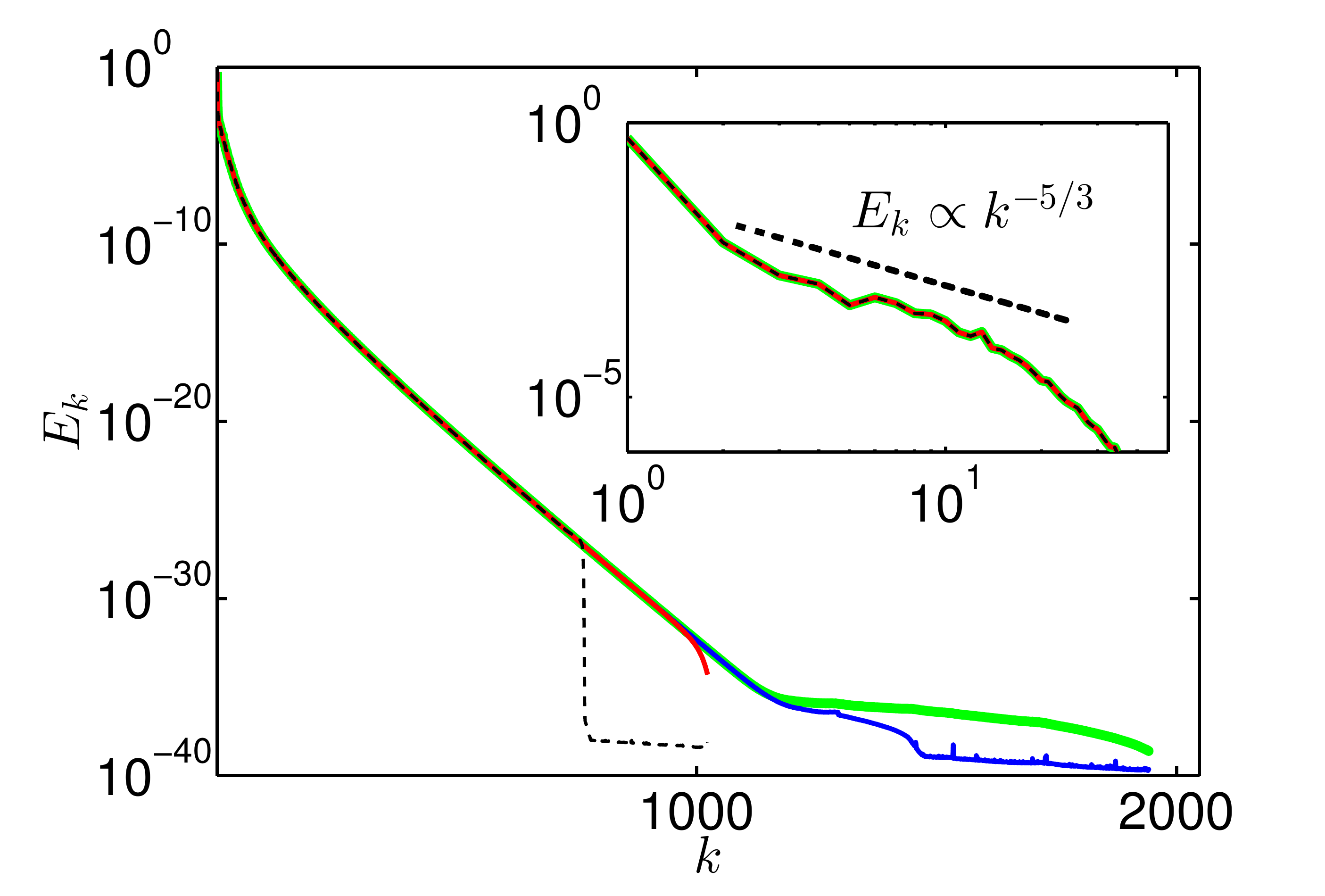}

\caption{{\it (Color on-line)} 
Energy spectrum $E_{k}(t)$ at $t=6$ for four simulations of $I_{1}$ initial condition from~\cite{agafontsev2015}: $2048^{3}$ total number of nodes with Hou \& Li filtering~(\ref{HLdumping}) (thick green), $2048^{3}$ nodes with $2/3$-dealiasing rule~(\ref{23dumping}) (blue), $1024^{3}$ nodes with Hou \& Li filtering~(\ref{HLdumping}) (thin red) and $1024^{3}$ nodes with $2/3$-dealiasing rule~(\ref{23dumping}) (dashed black). 
}
\label{fig:fig1}
\end{figure}

A careful analysis with different numerical algorithms was done to rule our the possibility, that the $5/3$ interval in the energy spectrum is influenced by the Fourier filtering~(\ref{HLdumping}). 
First, both the energy $E=(1/2)\int \mathbf{v}^2\,d^{3}\mathbf{r}$ and helicity $\Omega=\int (\mathbf{v}\cdot\omegabold)\,d^{3}\mathbf{r}$ are conserved in our simulations, with the relative error of $10^{-11}$ order. 
Second, the same energy spectrum is obtained with the cut-off function corresponding to the standard $2/3$-dealiasing rule 
\begin{eqnarray}\label{23dumping}
\rho(\mathbf{k})=\left\{ \begin{array}{rlllc} 
1 & \mbox{if} & \forall j: |k_{j}|\le 2K_{\max}^{(j)}/3, 
\\ 0 & \mbox{if} & \exists j: |k_{j}| > 2K_{\max}^{(j)}/3, \end{array}\right.
\end{eqnarray}
where $j=x,y,z$. 
Third, simulations with different grids perfectly converge. 
The latter two facts are illustrated in Fig.~\ref{fig:fig1}, which shows the energy spectrum at $t=6$ for four simulations of $I_{1}$ initial condition from~\cite{agafontsev2015}. 
The two simulations with Fourier filters~(\ref{HLdumping}) and~(\ref{23dumping}), both limited by $2048^{3}$ nodes, reach the final grid $972\times 2048\times 4096$ only at $t=6.08$. 
This means that at $t=6$ the aliasing and the filtering have negligible influence on the ``signal-containing'' harmonics for Hou \& Li filtering~(\ref{HLdumping}), and identically no influence for dealiasing rule filtering~(\ref{23dumping}). 
Nevertheless, the energy spectrum for these simulations practically coincides with that for other two simulations, with the same Fourier filters and limited by $1024^{3}$ nodes, which are affected at $t=6$ by both the aliasing and the filtering. 
All four simulations demonstrate the identical $5/3$-interval in the energy spectrum, and at $t=6$ have relative point-by-point difference $|\omegabold^{(1)}(\mathbf{r})-\omegabold^{(2)}(\mathbf{r})|/|\omegabold^{(1)}(\mathbf{r})|$ between any two of them $\omegabold^{(1)}(\mathbf{r})$ and $\omegabold^{(2)}(\mathbf{r})$ below $10^{-8}$, with the vorticity changing in the range $0.22\le|\omegabold(\mathbf{r})|\le 7.6$. 
Thus, one can conclude that the emergence of the interval in the energy spectrum with the Kolmogorov-like scaling is not affected by the specifics of our numerical scheme and should be related to fluid dynamics governed by the Euler equations. 


\section{Results}

We examine the local geometry of a high-vorticity structure using the Hessian matrix $\partial_{i}\partial_{j}|\omegabold|$ of second derivatives of vorticity modulus $|\omegabold|$ with respect to $(x,y,z)$, computed at the local vorticity maximum. 
The normal direction to such a structure is defined as the eigenvector corresponding to the largest of the three eigenvalues $|\lambda_{1}|\ge |\lambda_{2}|\ge |\lambda_{3}|$ of the Hessian. 
The characteristic scales are estimated with the local second-order approximation as $\ell_{i}=\sqrt{2\,\omega_{\max}/|\lambda_{i}|}$. 
Note that numerical determination of the local vorticity maximums is a nontrivial problem due to specific geometry of the pancake structures; we refer to~\cite{agafontsev2015} for description of the methods.

\begin{figure}[t]
\centering
\includegraphics[width=8.5cm]{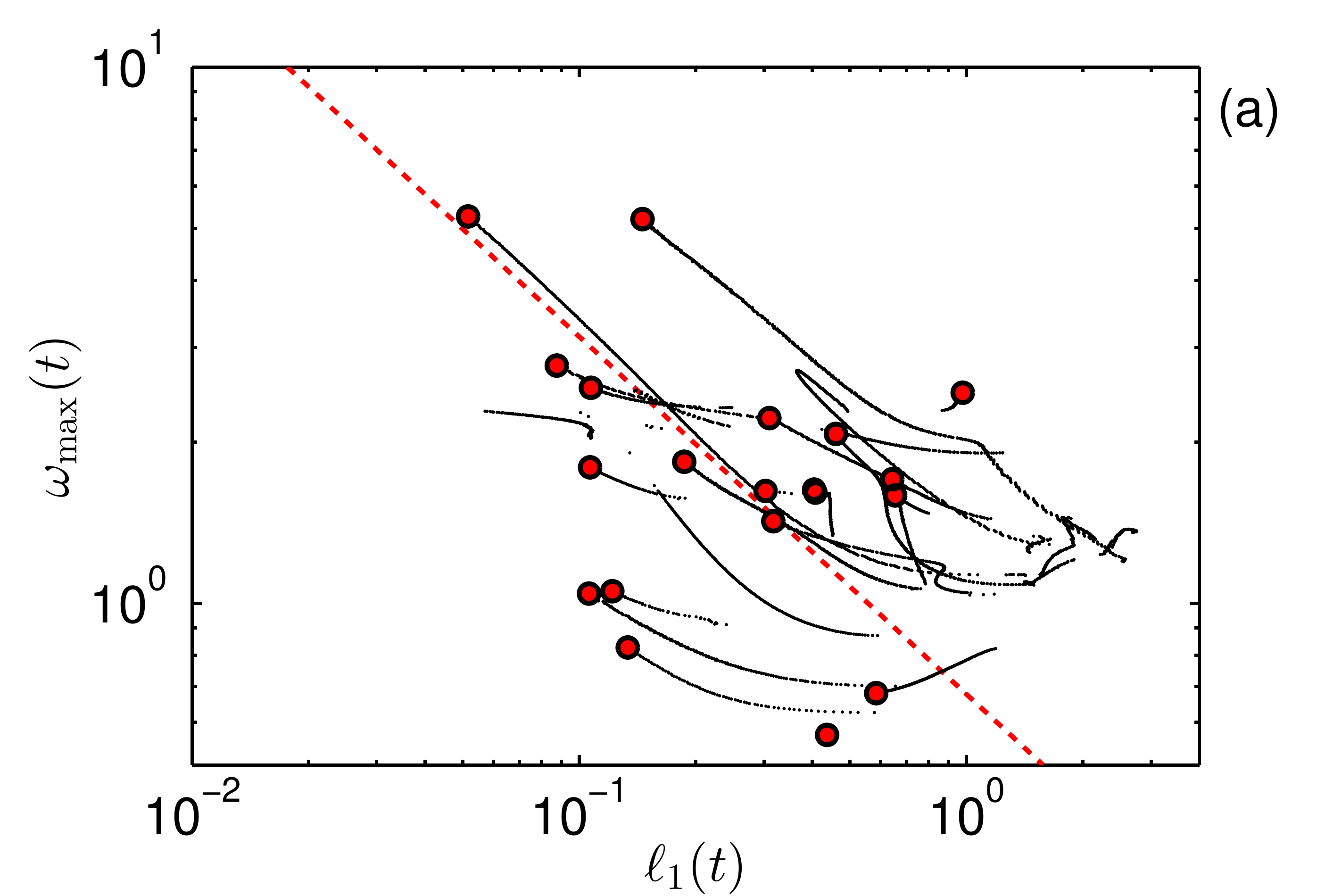}
\includegraphics[width=8.5cm]{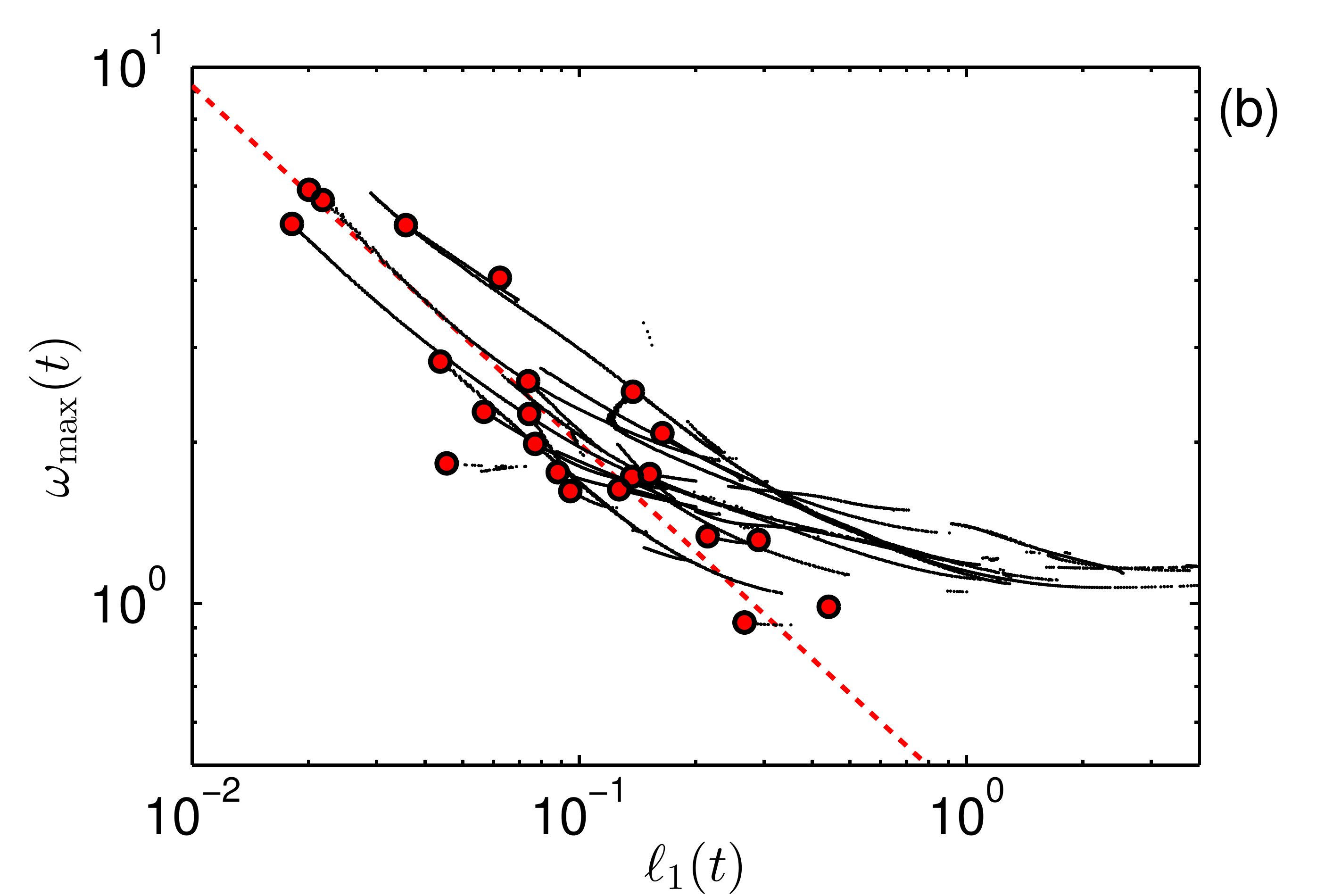}

\caption{{\it (Color on-line)} 
Local vorticity maximums $\omega_{\max}(t)$ vs. pancake thickness $\ell_{1}(t)$ during the evolution of the pancake structures for (a) $IC_{1}$ (Type I) and (b) $IC_{3}$ (Type III) initial conditions. 
Red circles mark local maximums at the final time, dashed red line indicates the power-law $\omega_{\max} \propto \ell_1^{-2/3}$. 
}
\label{fig:fig2}
\end{figure}

We observe that, for all 30 simulations, the regions of high vorticity represent pancake-like structures (thin vorticity sheets) of decreasing thickness $\ell_{1}$ and other two scales $\ell_{2}\sim\ell_{3}\sim 1$ not changing considerably. 
At the final simulation time, the span-to-thickness ratio reaches $\ell_{2}/\ell_{1}\sim \ell_{3}/\ell_{1}\sim 100$ for some of these structures. 
The pancakes develop in increasing number and demonstrate a clear tendency for asymptotic exponential vorticity growth $\omega_{\max}(t)\propto e^{t/T_{\omega}}$ and compression in the transversal direction $\ell_{1}(t)\propto e^{-t/T_{\ell}}$, with the relation $T_{\ell}/T_{\omega}\approx 2/3$. 
Here $\omega_{\max}(t)$ is the maximum of vorticity within the pancake. 
The $2/3$ relation leads to Kolmogorov-type scaling between the vorticity maximum and pancake thickness, 
\begin{equation}
\omega_{\max}(t)\propto \ell_{1}(t)^{-2/3},
\label{omega23}
\end{equation}
which is illustrated in Fig.~\ref{fig:fig2} on the examples of $IC_{1}$ (Type I) and $IC_{3}$ (Type III) initial conditions. 
Note that for some initial conditions and for some of the pancakes we observe deviations from scaling~(\ref{omega23}).
Thin pancake structures generate strongly anisotropic vorticity field in the Fourier space, concentrated in ``jets'' extended in the directions perpendicular to the pancakes. 
These jets occupy only a small fraction of the entire spectral space and provide the leading contribution to the energy spectrum~(\ref{energy_spectrum}). 
Thus, development of the pancake structures and evolution of the energy spectrum are directly related with each other. 
All these observations coincide with that, reported in our previous paper~\cite{agafontsev2015} on the basis of simulations of just two initial conditions. 

\begin{figure}[t]
\centering
\includegraphics[width=8.5cm]{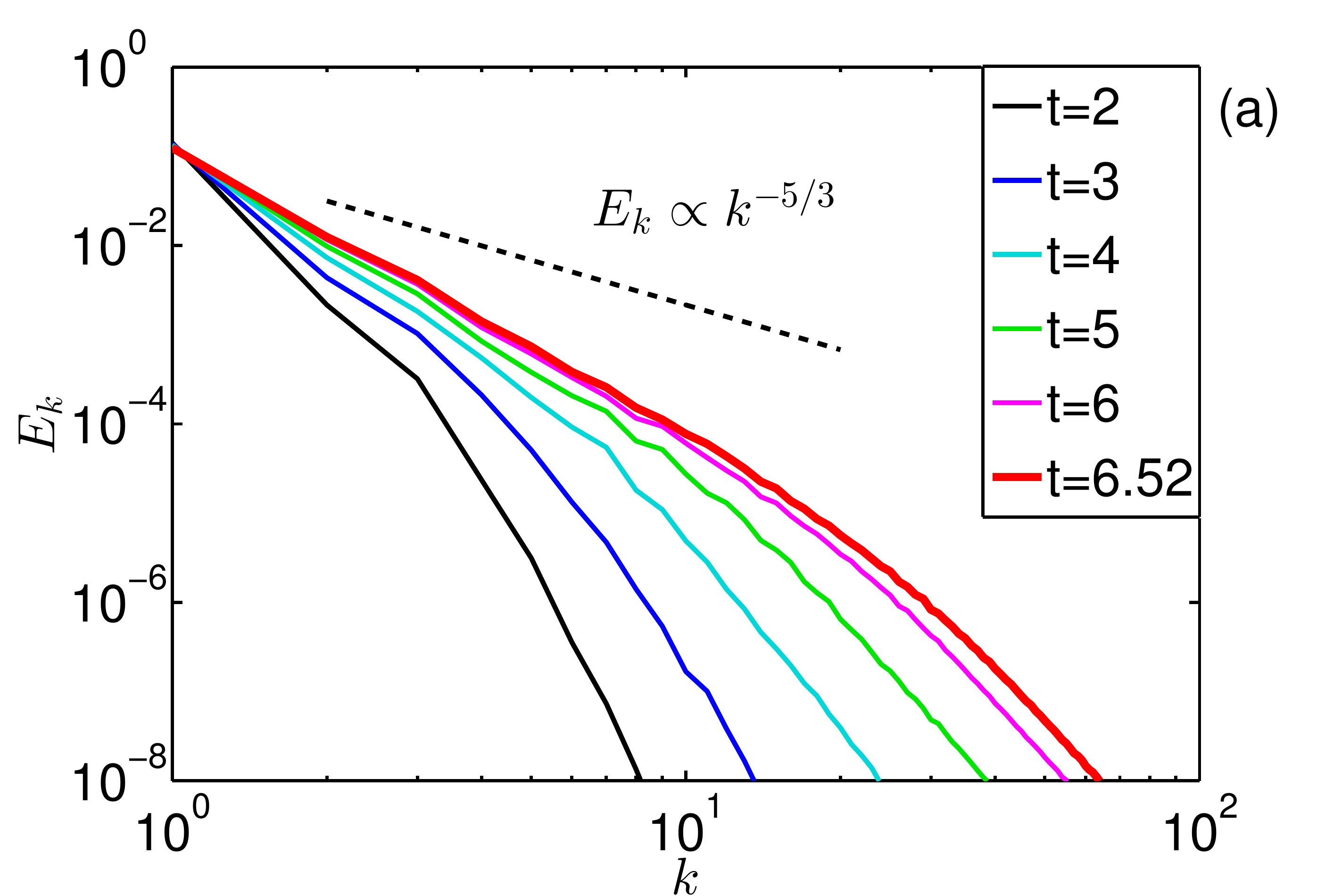}
\includegraphics[width=8.5cm]{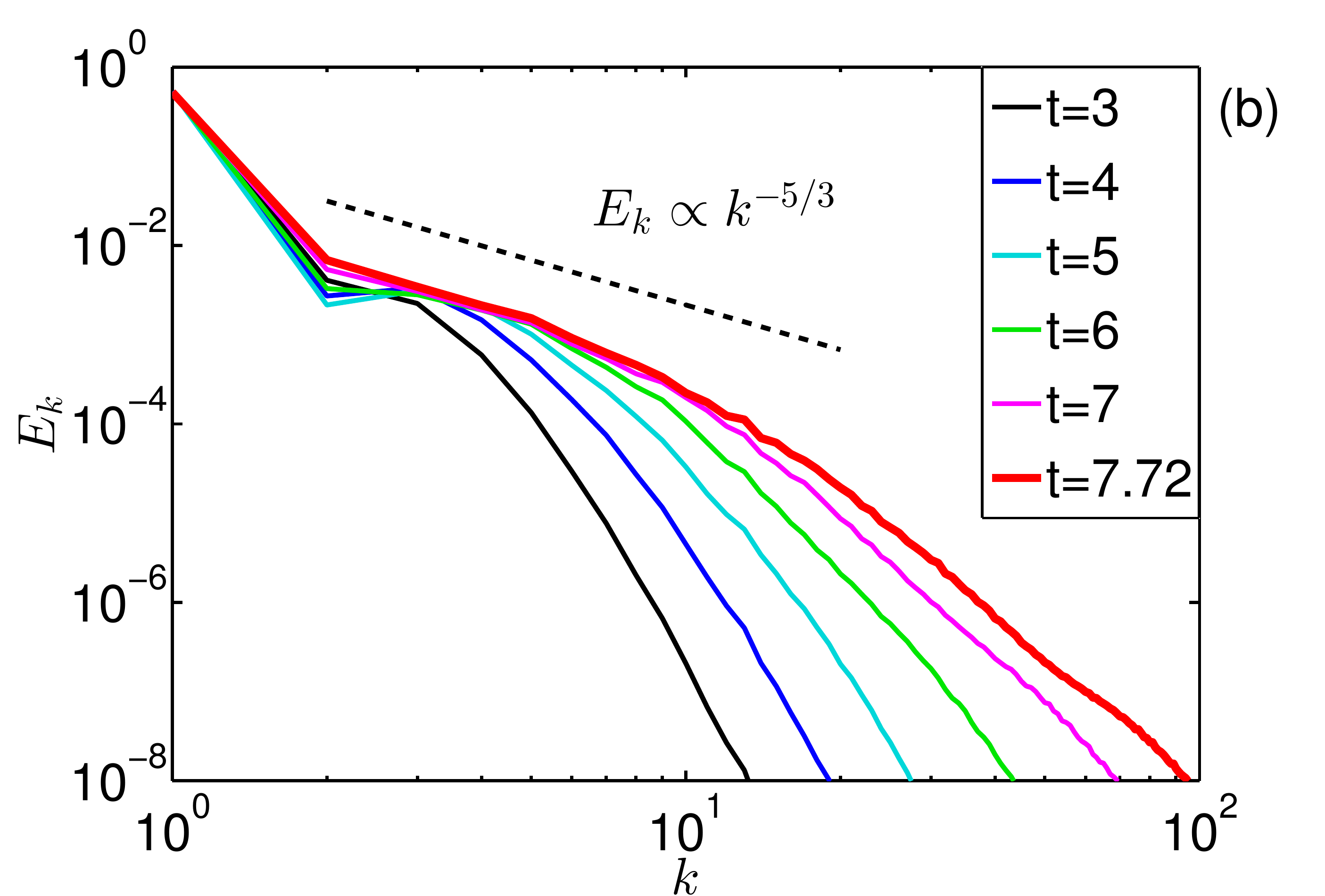}
\includegraphics[width=8.5cm]{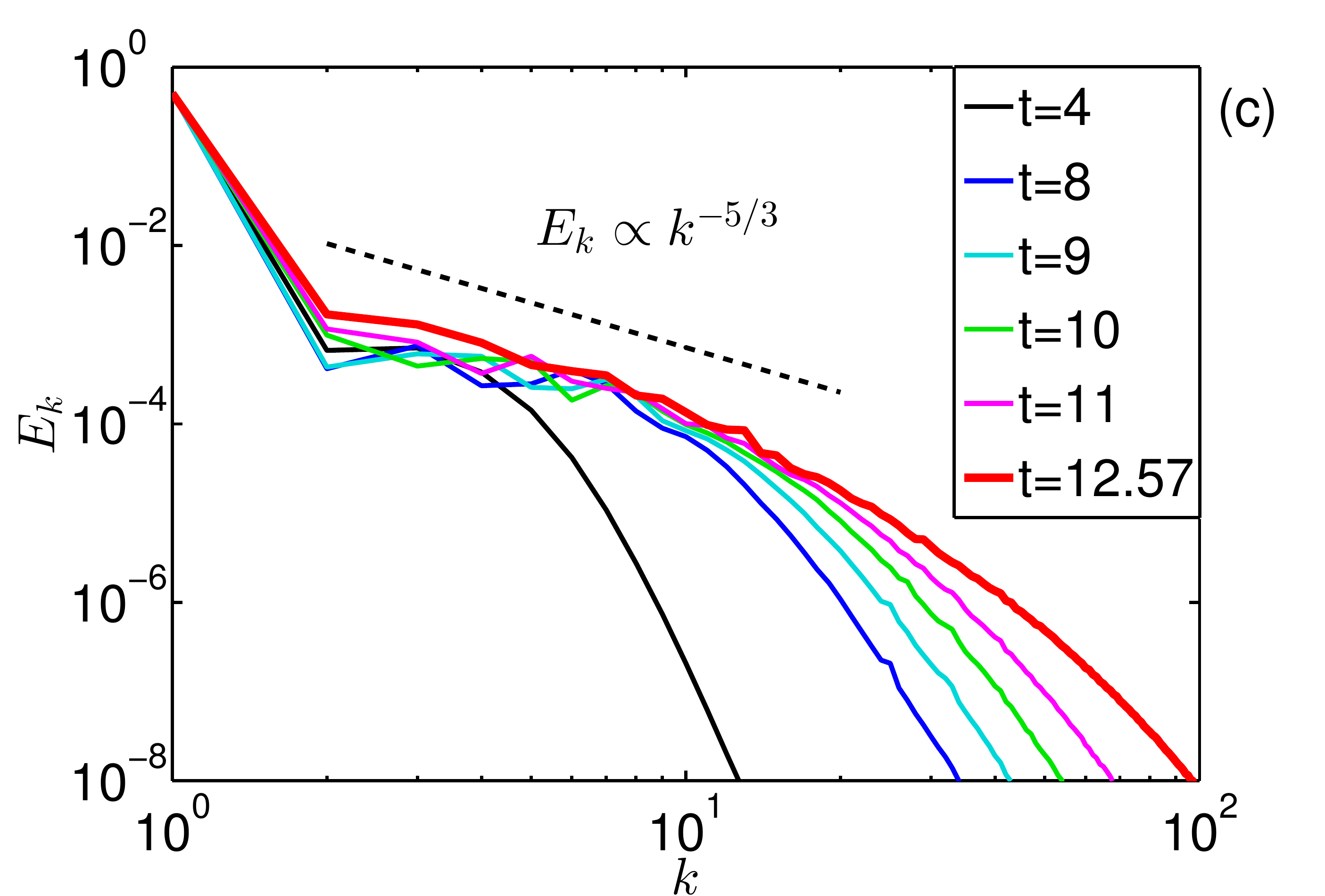}

\caption{{\it (Color on-line)} 
Energy spectrum $E_{k}(t)$ at different times for (a) $IC_{1}$ (Type I), (b) $IC_{2}$ (Type II) and (c) $IC_{3}$ (Type III) initial conditions. 
}
\label{fig:fig3}
\end{figure}

The pancake structures developed from Type I initial flows are oriented arbitrary. 
The corresponding simulations end in final grids similar to $1000^{3}$ with the increase of the global vorticity maximum by the factor $2.5-4$ compared with its initial value. 
Ten simulations of Type II initial conditions yield pancake structures with the normal directions typically within $45^{o}$ angle around $z$-axis, and end in grids similar to $800\times 800\times 1500$ with the global vorticity maximum increase by $4-6$ times.  
Ten simulations of Type III initial conditions end in grids similar to $600\times 600\times 2500$ with the global vorticity maximum increase by $3-5$ times. 
The pancakes have normal directions typically within $15^{o}$ angle to the $z$-axis. 

None of the ten simulations of Type I initial conditions develop a power-law interval in the energy spectrum with the exponent independent of time. 
This is illustrated in Fig.~\ref{fig:fig3}(a), where the energy spectrum behaves at small wavenumbers close to $E_{k}\propto k^{-5}$ at $t=3$, and close to $E_{k}\propto k^{-3}$ at the final time $t=6.52$. 
Five out of ten simulations of Type II initial conditions demonstrate the power-law interval with the scaling similar to the Kolmogorov law $E_{k}\propto k^{-5/3}$; the interval extends up to $2\lesssim k\lesssim 10$ for some of the simulations. 
Most of Type III initial flows develop sufficiently long power-law intervals up to $2\lesssim k\lesssim 20$ at the final time, with the scaling similar to the Kolmogorov one $E_{k}\propto k^{-5/3}$. 
The difference between the three types of initial conditions is illustrated in Fig.~\ref{fig:fig3}. 
Such a different behavior cannot be explained by different spatial resolution of the corresponding simulations: the $IC_{3}$ simulation reaches the stopping condition corresponding to $1000$ points in $z$-axis at $t\approx 11.5$, when the power-law interval in the energy spectrum is already developed up to $2\lesssim k\lesssim 20$. 
Note that for a few initial conditions we observe the power-law with a different exponent between $-8/3$ and $-4/3$; however, for most of the simulations this exponent is sufficiently close to $-5/3$. 


\section{Conclusions and discussions}

In this paper we examined the influence of initial conditions on the processes of pancake formation and the Kolmogorov energy spectrum development. 
With 30 new simulations, we systematically verified the key results of our previous study~\cite{agafontsev2015}, namely, that the regions of high vorticity represent exponentially compressing pancake-like structures, evolving with the scaling $\omega_{\max}(t)\propto\ell(t)^{-2/3}$ between the vorticity maximum and pancake thickness.
Collectively, these pancakes provide the leading contribution to the energy spectrum. 
Note that our paper~\cite{agafontsev2015} was based on just two simulations. 

The initial conditions for the present study were chosen as a combination of the shear flow and a random periodic flow. 
Such a mix allows one to influence the orientation of the emerging pancake structures, from fully random when the shear flow is absent to almost unidirectional when the perturbation is small. 
We observed that fully random initial flows did not develop the power-law interval in the energy spectrum with the exponent independent of time. 
On the contrary, most of the simulations of initial conditions taken as a mix of the shear flow with a small perturbation developed sufficiently long power-law intervals with the scaling similar to the Kolmogorov one $E_{k}\propto k^{-5/3}$. 

The results of the present paper demonstrate clearly that the development of the Kolmogorov energy spectrum can be influenced by the appropriate choice of initial conditions. 
The nature of this influence is yet to be discovered. 
The evolution of the energy spectrum is directly related to the development of the high-vorticity pancake structures. 
Then, different energy spectrum for simulations of different initial conditions may be related to distinctions in (1) composition of the pancake flows and (2) distribution of pancakes by scale, orientation and position in the physical space. 
In our next paper we are going to present a compelling evidence that the composition of the pancake flows is very similar for all pancakes emerging from all types of initial conditions. 
As for the distribution of pancakes by scale, orientation and position, an accurate numerical study in this direction implies the same spatial resolution for different initial conditions. 
For initially random periodic flows this means final grids similar to $2000^{3}$, for a considerable number of independent simulations. 
At the present moment, these demands exceed our numerical resources, and we plan to continue this study in the future. 
\\

{\it Acknowledgments.} 
Development of the numerical code and simulations were supported by the Russian Science Foundation (grant 14-22-00174), with the latter performed at the Novosibirsk Supercomputer Center (NSU). 
Analysis of the results was done at the Data Center of IMPA (Rio de Janeiro).
D.S.A. acknowledges the support from IMPA during the visits to Brazil. 
A.A.M. was supported by the CNPq (grant 302351/2015-9) and the Program FAPERJ Pensa Rio (grant E-26/210.874/2014).


%

\end{document}